\documentclass[journal]{IEEEtran}

\usepackage{cite}
\usepackage{amsmath,amssymb,amsfonts}
\usepackage{graphicx}
\usepackage{booktabs}
\usepackage{array}
\usepackage{tikz}
\usetikzlibrary{arrows.meta,positioning,fit,shapes.geometric}
\usepackage[colorlinks=false,hidelinks]{hyperref}

\graphicspath{{figures/}}

\newcommand{\Phalf}{\Phi_{1/2}}

\begin{document}

\title{Laser-Diode LiFi With Diffused-Beam Optics:\\
System-Level Modeling and a Cross-Validated ns-3 Simulation Framework}

\author{Hussain~Ahmad,
        Syed~Muhammad~Talha~Gillani,
        Toheed~Omer,
        and~Saleem~Aslam%
\thanks{H. Ahmad, S. M. T. Gillani, T. Omer, and S. Aslam are with the
Department of Electrical Engineering, Bahria University, Islamabad,
Pakistan (e-mail: hussainbuic@gmail.com).}%
\thanks{Simulation code (ns-3 module and Python link-level engine) is
available with this article.}}

\markboth{Journal of Lightwave Technology}%
{Ahmad \MakeLowercase{\textit{et al.}}: Laser-Diode LiFi With Diffused-Beam Optics}

\maketitle

\begin{abstract}
Laser diodes (LDs) promise an order-of-magnitude bandwidth advantage over
light-emitting diodes for indoor optical wireless access, but reported
prototype studies frequently leave the gap between hardware demonstrations
and system-level performance unquantified. This paper develops a complete,
reproducible system model of a diffused-beam LD LiFi transceiver---a 500-mW
laser source beam-shaped by a holographic diffuser, an intensity-modulation/
direct-detection (IM/DD) receiver, and adaptive $M$-QAM signaling---and
embeds it in two cross-validated simulators: an open ns-3 module providing
full-stack network simulation (channel, PHY, ARQ MAC, \texttt{NetDevice},
IP/UDP/TCP) and a Python link-level engine used for Monte-Carlo validation
of all analytical error models. Starting from a hardware prototype that
transferred data, real-time voice, and images over a 14-m line-of-sight
link, we identify and close the technical gaps typical of prototype-class
reports: serial-interface throughput ceilings misread as optical-link
capacity, absent noise modeling, unmeasurable error floors, and unexamined
beam-width/coverage trade-offs. The framework shows that the same optical
front end, freed of its 2-Mbaud UART bottleneck and driven at its 250-MHz
electrical bandwidth, supports 930~Mb/s net at 14~m under a $3.8\times
10^{-3}$ HD-FEC threshold with 16-QAM, scales to 1.86~Gb/s at 5~m with
256-QAM, and sustains on-off keying to 23.3~m; a $20^{\circ}$ diffuser
covers a $4.2$-m-radius cell of a standard room at desk height. Network
simulations over the ns-3 stack yield saturation goodput within 7\% of the
PHY line rate and sub-0.11-ms 99th-percentile latency at 70\% load. All
models, code, and figures are released for reproduction.
\end{abstract}


\IEEEpeerreviewmaketitle

\section{Introduction}

\IEEEPARstart{T}{he} radio spectrum below 300~GHz is a congested, regulated
and increasingly contested resource, while the optical spectrum offers
roughly $10^4$ times more bandwidth, is license-free, and is naturally
confined by opaque boundaries---an attractive property for both spatial
reuse and physical-layer security~\cite{chowdhury2018comparative,
pathak2015visible,haas2016what}. LiFi extends visible light communication
(VLC) from a point-to-point physical layer into a networked, multi-user
wireless access technology~\cite{haas2016what,islim2016modulation}.

Within optical wireless, the choice of emitter fundamentally constrains the
achievable rate. Phosphor-converted white LEDs are inexpensive and
eye-safe, but their 3-dB electrical bandwidth of a few to a few tens of MHz
limits single-emitter rates and forces heavy spectral-efficiency
engineering~\cite{karunatilaka2015led,tsonev2014gbs,islim2017towards}.
Laser diodes (LDs), by contrast, exhibit modulation bandwidths in the
hundreds of MHz to GHz range and much higher electrical-to-optical
conversion linearity at high current densities; GaN LDs have carried
9~Gb/s over a single 450-nm channel~\cite{chi2015gan}, white-light LD
lamps have delivered multi-Gb/s illumination-grade
links~\cite{wu2018white,chi2017violet}, and LD-based underwater links have
reached 1.5~Gb/s over 20~m~\cite{shen2016underwater,shen2017beyond}. The
principal objections to LDs---speckle, eye safety, and narrow
beams---are addressed by diffusive beam shaping: a holographic diffuser
converts the coherent beam into an extended, eye-safer Lambertian-like
source with a designer-chosen half-angle~\cite{zafar2017laser}.

This paper builds on a working LD-LiFi hardware prototype developed by the
authors: a 500-mW laser source switched by an N-channel MOSFET driver fed
from a CH340 USB--TTL serial bridge, beam-shaped by a holographic diffuser,
and received by a photodiode/laser-sensor front end, with modulation and
demodulation performed in MATLAB. The prototype transferred arbitrary data,
real-time voice, and $69\times 71$-pixel images over a 14-m indoor
line-of-sight (LOS) link. Such prototype-class studies---common in the
LiFi literature---demonstrate feasibility but typically leave four
technical gaps that prevent their results from being extrapolated or
reproduced:

\begin{enumerate}
\item \emph{Interface-limited throughput is conflated with link capacity.}
A serial bridge capped at 2~Mbaud bounds goodput at 1.6~Mb/s regardless of
the optics; timing-based rate estimates taken at the application layer can
exceed this by orders of magnitude and must not be attributed to the
optical channel.
\item \emph{No noise model.} Without shot- and thermal-noise budgets, the
observation of ``error-free'' operation at one distance carries no
information about margins, achievable modulation orders, or scaling.
\item \emph{No error-rate methodology.} Short qualitative tests cannot
resolve bit-error rates (BERs) at or below forward-error-correction (FEC)
thresholds; a BER-versus-distance characterization requires either long
measurements or a validated model.
\item \emph{Unexplored design space.} The diffuser half-angle trades peak
SNR against coverage; transmit power trades range against eye safety;
modulation order trades rate against reach. None of these trade-offs can
be quantified from a single hardware operating point.
\end{enumerate}

\subsection{Contributions}

We close these gaps by lifting the prototype into a complete, validated
system model and simulation framework:

\begin{itemize}
\item \textbf{Link model.} A closed-form IM/DD link budget for
diffused-beam LD LiFi: generalized-Lambertian LOS channel parameterized by
the diffuser semi-angle, optical filtering and non-imaging concentration,
shot plus TIA thermal noise, and Gray-mapped $M$-QAM/OOK error models
against a 7\%-overhead hard-decision FEC threshold
(Sections~\ref{sec:channel}--\ref{sec:modulation}).
\item \textbf{An open ns-3 module.} \texttt{lifi} implements the channel,
PHY, stop-and-wait ARQ MAC, and \texttt{NetDevice} abstractions so that
unmodified IPv4/UDP/TCP stacks, applications, and \texttt{FlowMonitor}
instrumentation run over the optical link; mobility-model-driven geometry
makes coverage and mobility studies native (Section~\ref{sec:simulator}).
\item \textbf{Cross-validation.} A Python link-level engine implements the
identical equations and validates every analytical BER curve by Monte-Carlo
simulation; the ns-3 test suite asserts numerical agreement between the two
implementations to 0.1\% (Section~\ref{sec:simulator}).
\item \textbf{Quantified design space.} BER/SNR/rate versus distance,
power, and diffuser angle; room-scale coverage maps; image-transfer
fidelity; and full-stack goodput/latency (Section~\ref{sec:results}).
\item \textbf{Prototype forensics.} A quantitative account of what the
hardware prototype did and did not establish, including the UART ceiling
analysis that reconciles application-layer rate estimates with the
physical link (Section~\ref{sec:loopholes}).
\end{itemize}

\section{Related Work}\label{sec:related}

Kahn and Barry's treatment of indoor infrared links established the
generalized-Lambertian LOS model and IM/DD noise analysis that underpin
modern VLC link budgets~\cite{kahn1997wireless}; Komine and Nakagawa
adapted it to white-LED lighting~\cite{komine2004fundamental}, and
Ghassemlooy \emph{et al.} give a comprehensive modeling
treatment~\cite{ghassemlooy2019optical}. Surveys position LiFi within
optical wireless technologies~\cite{chowdhury2018comparative,
pathak2015visible,karunatilaka2015led}, and IEEE~802.15.7 standardizes
short-range optical PHY/MAC layers~\cite{ieee802157}.

On the emitter side, LED links progressed from 3~Gb/s single
$\mu$LED~\cite{tsonev2014gbs} to 10~Gb/s-class violet
$\mu$LED OFDM~\cite{islim2017towards}, while LD links demonstrated
9~Gb/s QAM-OFDM at 450~nm~\cite{chi2015gan}, illumination-grade white
lasing links~\cite{wu2018white,chi2017violet}, real-time Gb/s LiFi
transceivers~\cite{gao2018real}, and long-reach underwater
links~\cite{shen2016underwater,shen2017beyond}. Zafar \emph{et al.} argue
the LD roadmap for gigabit-class VLC~\cite{zafar2017laser}. Ray-traced
indoor studies quantify realistic mobile multi-gigabit
environments~\cite{hussein2015mobile}, receiver-side innovations include
$\mu$-photodetectors~\cite{ho2018gigabit} and organic detectors with
neural equalization~\cite{ghassemlooy2013visible}, and multi-user beam
control has been demonstrated with compound-eye
transmitters~\cite{cogalan2015power}.

At the network level, ns-3~\cite{riley2010ns3} is the reference
discrete-event simulator for wireless protocol research, and VLC modules
for LED links have been proposed~\cite{aldalbahi2016ns3}. To our
knowledge, no published ns-3 module models \emph{laser-diode} LiFi with
diffuser-parameterized beam shaping, nor pairs the network simulator with
an independently implemented, Monte-Carlo-validated link engine. This
paper fills that gap and anchors the model to a physical prototype.

\section{System Architecture}\label{sec:architecture}

\begin{figure*}[!t]
\centering
\resizebox{\textwidth}{!}{%
\begin{tikzpicture}[
  node distance=3.2mm and 4.5mm,
  blk/.style={draw, rounded corners=1pt, minimum height=6.5mm,
              minimum width=15mm, align=center, font=\scriptsize},
  opt/.style={blk, fill=blue!8},
  ele/.style={blk, fill=orange!10},
  net/.style={blk, fill=green!8},
  lbl/.style={font=\scriptsize\itshape},
  arr/.style={-{Stealth[length=1.6mm]}, semithick}]
\node[net] (src) {Data / voice /\\ image source};
\node[ele, right=of src] (mod) {Modulation\\ (OOK--256-QAM)};
\node[ele, right=of mod] (drv) {MOSFET\\ LD driver};
\node[opt, right=of drv] (ld) {500-mW\\ laser diode};
\node[opt, right=of ld] (dif) {Holographic\\ diffuser $\Phalf$};
\node[blk, right=of dif, fill=gray!12, minimum width=17mm]
  (ch) {LOS Lambertian\\ channel $H(0)$};
\node[opt, right=of ch] (pd) {Photodiode\\ $+$ concentrator};
\node[ele, right=of pd] (tia) {TIA\\ $R_f$};
\node[ele, right=of tia] (dem) {Demodulation\\ $+$ FEC};
\node[net, right=of dem] (snk) {Data\\ sink};
\draw[arr] (src) -- (mod);
\draw[arr] (mod) -- (drv);
\draw[arr] (drv) -- (ld);
\draw[arr] (ld) -- (dif);
\draw[arr, dashed] (dif) -- (ch);
\draw[arr, dashed] (ch) -- (pd);
\draw[arr] (pd) -- (tia);
\draw[arr] (tia) -- (dem);
\draw[arr] (dem) -- (snk);
\node[lbl, below=1.2mm of drv] {electrical};
\node[lbl, below=1.2mm of ch] {optical, free space};
\node[lbl, below=1.2mm of tia] {electrical};
\end{tikzpicture}%
}
\caption{Diffused-beam LD LiFi transceiver architecture. In the hardware
prototype the modulation/demodulation blocks run in MATLAB behind a CH340
USB--TTL bridge; in the simulation framework they are the \texttt{LifiPhy}
abstraction driven at the LD's full electrical bandwidth.}
\label{fig:arch}
\end{figure*}
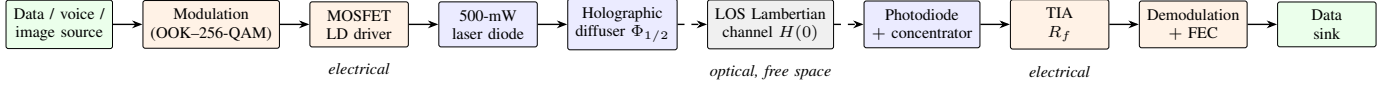

Fig.~\ref{fig:arch} shows the system. A data source feeds a digital
modulator; the resulting intensity waveform switches the LD through a
MOSFET driver. The coherent beam passes a holographic diffuser that (i)
destroys spatial coherence, easing eye-safety constraints of the extended
source~\cite{iec60825,zafar2017laser}, and (ii) shapes the emission into
an approximately Lambertian lobe of order $m$ set by the semi-angle at
half power $\Phalf$. At the receiver, a non-imaging concentrator and
optical filter collect light onto a photodiode; a transimpedance
amplifier (TIA) converts photocurrent to voltage; the demodulator
recovers bits, protected by hard-decision FEC.

The hardware prototype realized this chain at reduced scale: a CH340
USB--TTL bridge (maximum 2~Mbaud) drove the MOSFET gate, MATLAB performed
(de)modulation offline, and a laser-sensor module served as receiver.
The prototype verified end-to-end data, real-time voice, and image
transfer at 14~m indoors. Section~\ref{sec:loopholes} quantifies exactly
which performance claims that hardware can and cannot support.

\section{Optical Channel and Receiver Model}\label{sec:channel}

\subsection{LOS Channel Gain}

The diffused LD emission is modeled as a generalized Lambertian source of
order
\begin{equation}
m = -\frac{\ln 2}{\ln\cos\Phalf},
\label{eq:lambertian-order}
\end{equation}
so that $\Phalf=60^{\circ}$ gives $m=1$ (ideal Lambertian) while
$\Phalf=20^{\circ}$ gives $m=11.14$, concentrating power near boresight.
For a receiver of physical area $A$ at distance $d$, emission angle
$\phi$ and incidence angle $\psi$, the DC channel gain
is~\cite{kahn1997wireless,komine2004fundamental}
\begin{equation}
H(0)=
\begin{cases}
\dfrac{(m+1)A}{2\pi d^{2}}\cos^{m}\!\phi\; T_s\, g(\psi)\cos\psi,
 & 0\le\psi\le\Psi_c,\\[2mm]
0, & \psi>\Psi_c,
\end{cases}
\label{eq:H0}
\end{equation}
where $T_s$ is the optical filter transmission and the non-imaging
concentrator of refractive index $n$ and field-of-view semi-angle
$\Psi_c$ contributes
\begin{equation}
g(\psi)=\frac{n^{2}}{\sin^{2}\Psi_c},\qquad 0\le\psi\le\Psi_c .
\label{eq:conc}
\end{equation}
The received optical power is $P_r=H(0)\,P_t$.

\subsection{Receiver Noise and SNR}

Direct detection converts $P_r$ into photocurrent $I_{ph}=\mathcal{R}P_r$
with responsivity $\mathcal{R}$. Over receiver bandwidth $B$, shot noise
(including a background-light current $I_{bg}$) and TIA thermal noise
contribute
\begin{align}
\sigma_{\mathrm{sh}}^{2} &= 2q\,(I_{ph}+I_{bg})\,B,
\label{eq:shot}\\
\sigma_{\mathrm{th}}^{2} &= \frac{4k_BT}{R_f}\,B,
\label{eq:thermal}
\end{align}
yielding the electrical SNR
\begin{equation}
\gamma=\frac{(\mathcal{R}P_r)^{2}}
{\sigma_{\mathrm{sh}}^{2}+\sigma_{\mathrm{th}}^{2}}.
\label{eq:snr}
\end{equation}

Table~\ref{tab:params} lists the parameter set, chosen to match the
prototype's components scaled to the LD's full electrical bandwidth.
With these values the link is background-shot-noise limited:
at 14~m and $\Phalf=20^{\circ}$, $P_r=1.13$~$\mu$W and
$\gamma=17.3$~dB.

\begin{table}[!t]
\caption{System Parameters (Prototype-Derived Defaults)}
\label{tab:params}
\centering
\begin{tabular}{lll}
\toprule
Symbol & Meaning & Value\\
\midrule
$P_t$ & transmit optical power & 500 mW\\
$\Phalf$ & diffuser semi-angle & $20^{\circ}$--$60^{\circ}$\\
$A$ & photodetector area & 1 cm$^{2}$\\
$\mathcal{R}$ & responsivity & 0.6 A/W\\
$T_s$ & filter transmission & 0.9\\
$n$ & concentrator index & 1.5\\
$\Psi_c$ & receiver field of view & $70^{\circ}$\\
$B$ & electrical bandwidth & 250 MHz\\
$I_{bg}$ & background photocurrent & 100 $\mu$A\\
$T$ & receiver temperature & 298 K\\
$R_f$ & TIA feedback resistance & 10 k$\Omega$\\
FEC & HD-FEC threshold / overhead & $3.8\times10^{-3}$ / 7\%\\
\bottomrule
\end{tabular}
\end{table}

\section{Modulation and Error Models}\label{sec:modulation}

The PHY supports OOK, BPSK, QPSK, and square 16-/64-/256-QAM on an
electrical subcarrier (DC-biased for unipolarity), all at symbol rate
$B$. With Gray mapping, the standard electrical-domain error
probabilities apply~\cite{ghassemlooy2019optical}: for OOK
$P_b=Q(\sqrt{\gamma})$, for BPSK $P_b=Q(\sqrt{2\gamma_b})$, and for
square $M$-QAM
\begin{equation}
P_b \approx \frac{4}{\log_2 M}\Bigl(1-\tfrac{1}{\sqrt M}\Bigr)
Q\!\left(\sqrt{\frac{3\gamma}{M-1}}\right).
\label{eq:mqam}
\end{equation}
A rate-$(1-\rho)$ hard-decision FEC with $\rho=7\%$ overhead corrects any
pre-FEC BER at or below $3.8\times10^{-3}$; the net rate of modulation
$M$ is then $R_{\mathrm{net}} = B\log_2 M\,(1-\rho)$ when
$P_b\le 3.8\times10^{-3}$ and the link is in outage otherwise. The
adaptive-rate policy selects the highest-order modulation meeting the
threshold.

\begin{figure}[!t]
\centering
\includegraphics[width=\columnwidth]{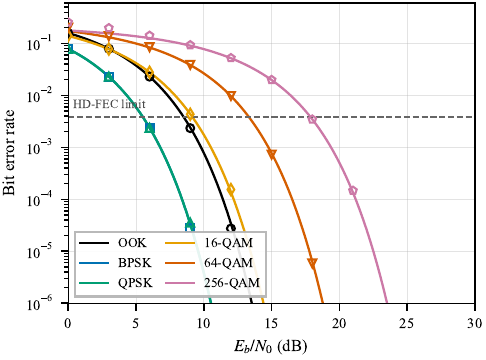}
\caption{BER versus $E_b/N_0$: analytical curves (lines) and Monte-Carlo
validation (open markers, $\ge 1.2\times10^{6}$ bits/point). BPSK and QPSK
coincide, as expected. The dashed line is the $3.8\times10^{-3}$ HD-FEC
threshold.}
\label{fig:bersnr}
\end{figure}

Every closed-form curve used in this paper is verified by Monte-Carlo
simulation of at least $1.2\times10^{6}$ bits per point in the Python
engine (Section~\ref{sec:simulator}); Fig.~\ref{fig:bersnr} overlays the
Monte-Carlo estimates (markers) on the analytical curves (lines).

\section{Simulation Framework}\label{sec:simulator}

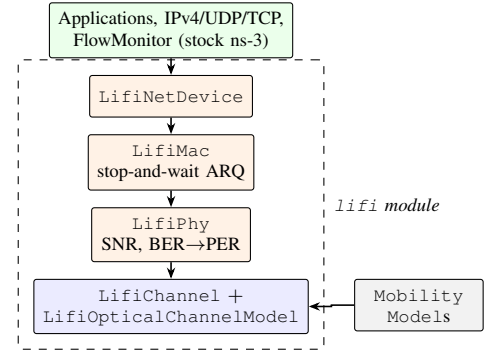
\begin{figure}[!t]
\centering
\begin{tikzpicture}[
  node distance=2.2mm and 3mm,
  cls/.style={draw, rounded corners=1pt, minimum height=5.5mm,
              minimum width=21mm, align=center, font=\scriptsize},
  grp/.style={draw, dashed, rounded corners=2pt, inner sep=2mm},
  arr/.style={-{Stealth[length=1.5mm]}, semithick}]
\node[cls, fill=green!8] (app) {Applications, IPv4/UDP/TCP,\\ FlowMonitor (stock ns-3)};
\node[cls, fill=orange!10, below=of app] (dev) {\texttt{LifiNetDevice}};
\node[cls, fill=orange!10, below=of dev] (mac) {\texttt{LifiMac}\\ stop-and-wait ARQ};
\node[cls, fill=orange!10, below=of mac] (phy) {\texttt{LifiPhy}\\ SNR, BER$\to$PER};
\node[cls, fill=blue!8, below=of phy] (chn) {\texttt{LifiChannel} $+$\\ \texttt{LifiOpticalChannelModel}};
\node[cls, fill=gray!10, right=6mm of chn.north east, anchor=north west,
      minimum width=17mm] (mob) {\texttt{Mobility}\\ \texttt{Model}s};
\draw[arr] (app) -- (dev);
\draw[arr] (dev) -- (mac);
\draw[arr] (mac) -- (phy);
\draw[arr] (phy) -- (chn);
\draw[arr] (mob.west) -- ++(-3mm,0) |- (chn.east);
\node[grp, fit=(dev)(mac)(phy)(chn), label={[font=\scriptsize\itshape]right:\texttt{lifi} module}] {};
\end{tikzpicture}
\caption{Architecture of the ns-3 \texttt{lifi} module. Geometry is read
live from standard ns-3 mobility models, so coverage and mobility studies
use stock ns-3 tooling.}
\label{fig:ns3arch}
\end{figure}

\subsection{ns-3 Module}

Fig.~\ref{fig:ns3arch} shows the module. \texttt{LifiOpticalChannelModel}
evaluates \eqref{eq:lambertian-order}--\eqref{eq:conc} from the live
positions and orientations of transmitter and receiver;
\texttt{LifiChannel} delivers each frame to all attached PHYs with
propagation delay $d/c$. \texttt{LifiPhy} computes
\eqref{eq:shot}--\eqref{eq:snr} per reception, maps SNR to BER via the
modulation catalogue, converts to packet error probability
$\mathrm{PER}=1-(1-P_b)^{8L}$ for an $L$-byte frame, and draws a Bernoulli
outcome; overlapping receptions are treated as collisions.
\texttt{LifiMac} implements stop-and-wait ARQ with bounded retries;
\texttt{LifiNetDevice} exposes the link as a standard \texttt{NetDevice},
so IP stacks, traffic generators, and \texttt{FlowMonitor} run unmodified.
Four example programs reproduce every scenario in this paper as CSV
output.

\subsection{Cross-Validated Link Engine}

An independent Python implementation of
\eqref{eq:lambertian-order}--\eqref{eq:mqam} provides Monte-Carlo BER
validation, image-transfer experiments, and the network-level queueing
model used for pre-study of the ns-3 scenarios. The ns-3 unit-test suite
pins the C++ implementation to reference values exported from the Python
engine (channel gain, Lambertian order, SNR, BER thresholds) at 0.1\%
relative tolerance, guaranteeing that both simulators realize the same
physics. This two-implementation strategy catches transcription errors
that single-implementation studies cannot.

\section{Results}\label{sec:results}

\subsection{Link Budget Versus Distance and Beam Width}

\begin{figure}[!t]
\centering
\includegraphics[width=\columnwidth]{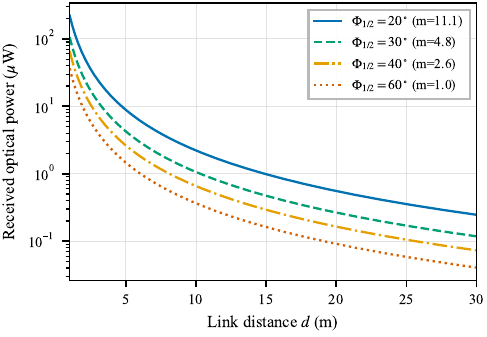}
\caption{Received optical power versus distance for four diffuser
semi-angles ($P_t=500$~mW). Narrower diffusion concentrates power near
boresight: at 14~m, $\Phalf=20^{\circ}$ delivers $1.13~\mu$W versus
$0.19~\mu$W for $60^{\circ}$.}
\label{fig:gain}
\end{figure}

\begin{figure}[!t]
\centering
\includegraphics[width=\columnwidth]{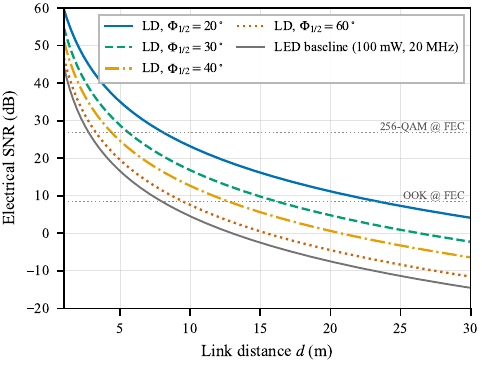}
\caption{Electrical SNR versus distance at $B=250$~MHz for the LD system
(four diffuser angles) and a representative LED baseline (100~mW,
$60^{\circ}$, 20~MHz). Dotted horizontals mark the SNRs required to reach
the FEC threshold with OOK (8.5~dB) and 256-QAM (26.9~dB).}
\label{fig:snrdist}
\end{figure}

Fig.~\ref{fig:gain} shows received power versus distance for
$\Phalf\in\{20^{\circ},30^{\circ},40^{\circ},60^{\circ}\}$, and
Fig.~\ref{fig:snrdist} the resulting SNR at full bandwidth. The
$1/d^{2}$ LOS law and the $(m+1)\cos^{m}\phi$ directivity together imply
that halving the diffusion angle buys roughly 6--8~dB of boresight SNR:
at 14~m the link clears the OOK FEC threshold by 8.8~dB with the
$20^{\circ}$ diffuser but falls 6.8~dB short with the $60^{\circ}$
diffuser. The LED baseline---representative of a 100-mW, 20-MHz
white-LED front end---remains usable only below 8~m even with adaptive
signaling, quantifying the LD advantage at equal detector hardware.

\subsection{Error Rate and Achievable Rate}

\begin{figure}[!t]
\centering
\includegraphics[width=\columnwidth]{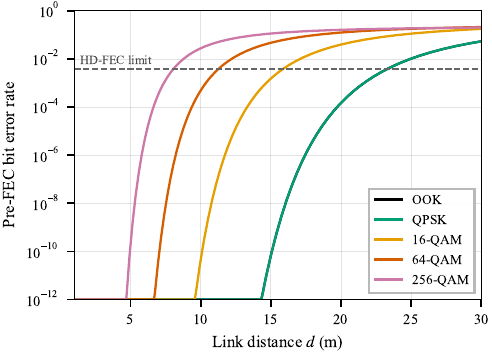}
\caption{Pre-FEC BER versus distance for each modulation
($\Phalf=20^{\circ}$, $B=250$~MHz). Crossings with the FEC threshold
define maximum reaches: 8.0~m (256-QAM), 11.2~m (64-QAM), 15.9~m
(16-QAM), and 23.3~m (OOK/QPSK).}
\label{fig:berdist}
\end{figure}

\begin{figure}[!t]
\centering
\includegraphics[width=\columnwidth]{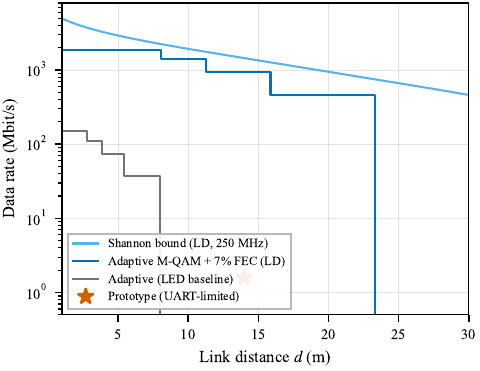}
\caption{Achievable rate versus distance: Shannon bound, adaptive
$M$-QAM with 7\% FEC overhead for the LD system and the LED baseline, and
the hardware prototype's UART-limited operating point (14~m, 1.6~Mb/s).}
\label{fig:ratedist}
\end{figure}

Fig.~\ref{fig:berdist} converts SNR to pre-FEC BER per modulation. The
adaptive-rate staircase (Fig.~\ref{fig:ratedist}) selects 256-QAM to
8.0~m (1.86~Gb/s net), 64-QAM to 11.2~m (1.40~Gb/s), 16-QAM to 15.9~m
(930~Mb/s), QPSK to 23.3~m (465~Mb/s), and OOK as the long-reach floor
(233~Mb/s). Two conclusions follow. First, the prototype's 14-m
demonstration sits comfortably inside the 16-QAM region: its optical
front end, driven at full bandwidth, supports \emph{930~Mb/s net} at the
demonstrated range---577$\times$ the UART-limited goodput actually
measurable on the hardware. Second, the gap between the staircase and
the Shannon bound (2.2--4.6~b/s/Hz at 5--20~m) bounds what
higher-order constellations, OFDM bit-loading, or soft-decision FEC
could still recover.

\subsection{Signal Quality and Bandwidth--Rate Trade}

\begin{figure*}[!t]
\centering
\includegraphics[width=\textwidth]{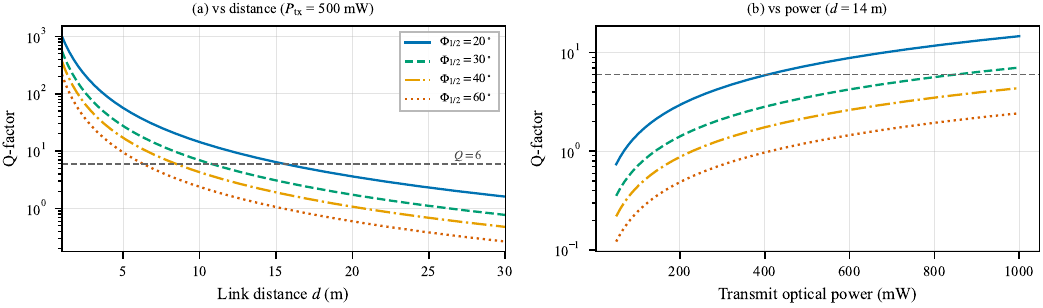}
\caption{OOK Q-factor (a) versus distance at $P_t=500$~mW and (b) versus
transmit power at $d=14$~m, for four diffuser angles. The $Q=6$ line
corresponds to BER $=10^{-9}$ uncoded.}
\label{fig:qfactor}
\end{figure*}

\begin{figure*}[!t]
\centering
\includegraphics[width=\textwidth]{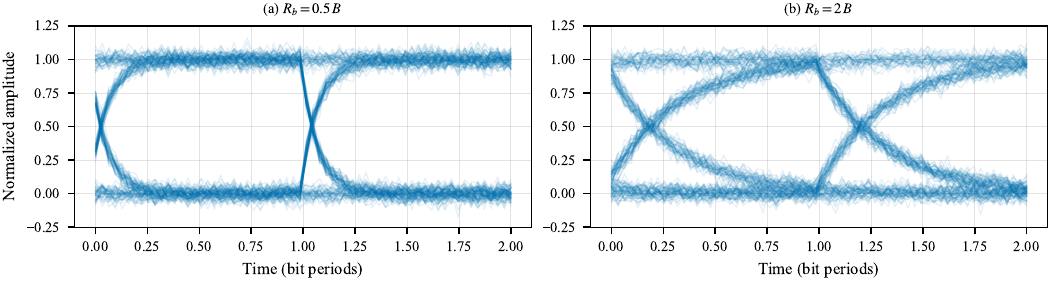}
\caption{Simulated OOK eye diagrams after a first-order receiver of
bandwidth $B$: (a) $R_b=0.5B$ leaves a fully open eye; (b) $R_b=2B$
closes it with intersymbol interference---the regime a
bandlimited front end enters when pushed past its electrical bandwidth.}
\label{fig:eye}
\end{figure*}

The Q-factor analysis (Fig.~\ref{fig:qfactor}) reproduces the thesis's
distance/power sweeps under the validated noise model: at 14~m with the
$20^{\circ}$ diffuser, $Q=7.4$, exceeding the $Q=6$ ($10^{-9}$) mark
without FEC; the $60^{\circ}$ diffuser requires either $\sim$4$\times$
the power or FEC assistance at the same range. Fig.~\ref{fig:eye}
illustrates the bandwidth mechanism behind the rate--reach trade: driving
a front end past its electrical bandwidth closes the eye through
intersymbol interference, which is why the framework caps the symbol rate
at $B$ rather than extrapolating rate from timing measurements.

\subsection{Coverage}

\begin{figure*}[!t]
\centering
\includegraphics[width=\textwidth]{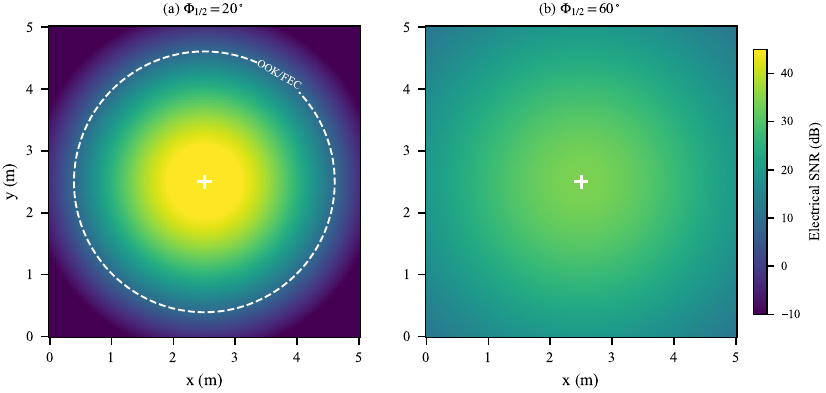}
\caption{Desk-height ($0.85$~m) SNR maps of a $5\times5\times3$-m room
with a ceiling-mounted downward transmitter: (a) $\Phalf=20^{\circ}$
yields a high-SNR cell with the OOK/FEC boundary (dashed) at
$\approx$2.1-m radius; (b) $\Phalf=60^{\circ}$ trades peak SNR for
wall-to-wall moderate SNR.}
\label{fig:coverage}
\end{figure*}

Fig.~\ref{fig:coverage} maps SNR over the receiver plane of a standard
room. The narrow diffuser produces a cell with $>40$~dB SNR at the
center---supporting 256-QAM---while the wide diffuser floods the room
with 15--25~dB, supporting OOK/QPSK everywhere but higher orders only
near the center. This is the quantitative form of the LiFi attocell
design choice: dense high-rate cells versus uniform moderate-rate
illumination~\cite{haas2016what,cogalan2015power}.

\subsection{The Prototype's Interface Ceiling}\label{sec:uart}

\begin{figure}[!t]
\centering
\includegraphics[width=\columnwidth]{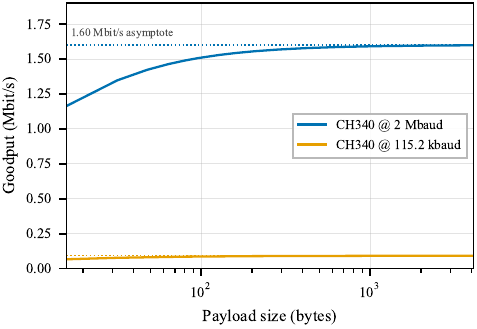}
\caption{Goodput of a CH340-class UART bridge versus payload size with
8N1 framing and 6~B/frame protocol overhead. At the prototype's maximum
2~Mbaud, goodput saturates at 1.59~Mb/s for 1500-B payloads---the true
ceiling of any measurement taken through this interface.}
\label{fig:uart}
\end{figure}

Fig.~\ref{fig:uart} quantifies the serial bridge: 8N1 framing costs 20\%
of the line rate, protocol overhead the rest, capping goodput at
1.59~Mb/s (1500-B payloads) at 2~Mbaud. Application-layer estimates of
the form $R=ND/t$---modulation-order-weighted payload over MATLAB-side
elapsed time---can exceed this by orders of magnitude because $t$
excludes serialization and buffering; the thesis's reported 127~Mb/s is
an estimate of this kind and should be read as a processing-rate figure,
not link throughput. The framework replaces it with the defensible chain:
measured 14-m error-free operation at UART rates $\Rightarrow$ validated
link model $\Rightarrow$ 930~Mb/s net at 250-MHz bandwidth
(Fig.~\ref{fig:ratedist}).

\subsection{Image Transfer Fidelity}

\begin{figure}[!t]
\centering
\includegraphics[width=\columnwidth]{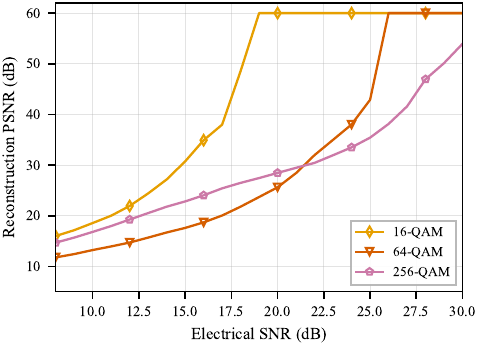}
\caption{Reconstruction PSNR of the $69\times71$ test image versus SNR
for three constellations (no FEC), simulated bit-exactly through the
Gray-mapped QAM chain.}
\label{fig:psnr}
\end{figure}

\begin{figure*}[!t]
\centering
\includegraphics[width=\textwidth]{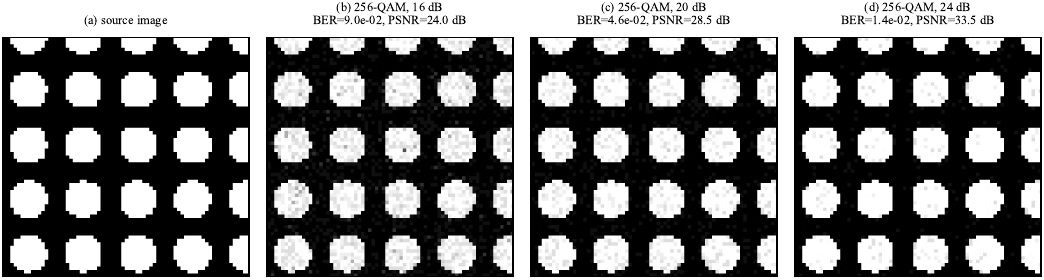}
\caption{Uncoded 256-QAM image transfer at three SNRs: (a) source; (b)
16~dB, BER $9.0\times10^{-2}$; (c) 20~dB, BER $4.6\times10^{-2}$;
(d) 24~dB, BER $1.4\times10^{-2}$. With the HD-FEC of
Section~\ref{sec:modulation}, all residual errors vanish above the
26.9-dB threshold.}
\label{fig:imagepanel}
\end{figure*}

The thesis's image experiment is reproduced bit-exactly
(Figs.~\ref{fig:psnr} and \ref{fig:imagepanel}): the $69\times71$ test
pattern is packed to bits, Gray-mapped, transmitted through the AWGN
electrical channel, and reassembled. PSNR rises $\sim$5~dB per 4~dB of
SNR until the error floor vanishes; with FEC, delivery is exact at every
operating point on the adaptive staircase, which is how the framework
turns the prototype's qualitative "image received" result into a
quantitative fidelity specification.

\subsection{Full-Stack Network Performance}

\begin{figure}[!t]
\centering
\includegraphics[width=\columnwidth]{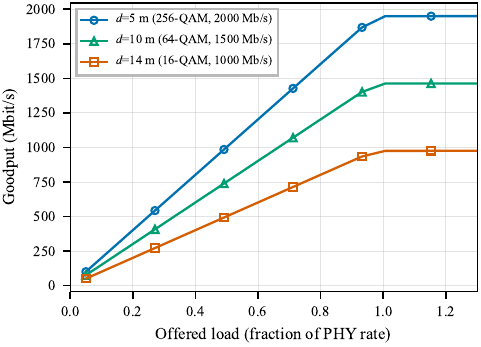}
\caption{Goodput versus offered load over the stop-and-wait ARQ link at
three distances (adaptive modulation; post-FEC residual BER $10^{-9}$).
Saturation reaches 93--96\% of the PHY line rate; the shortfall is ACK
serialization.}
\label{fig:load}
\end{figure}

\begin{figure}[!t]
\centering
\includegraphics[width=\columnwidth]{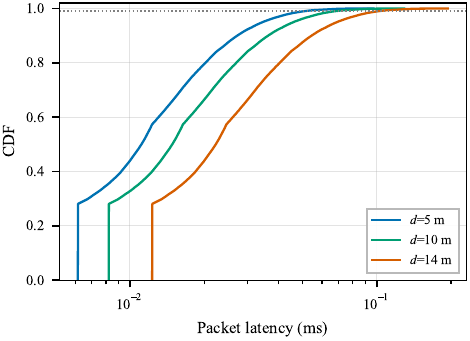}
\caption{Packet-latency CDFs at 70\% load. 99th-percentile latency is
51~$\mu$s at 5~m, 68~$\mu$s at 10~m, and 102~$\mu$s at 14~m,
tracking the PHY rate selected by the adaptive policy.}
\label{fig:latency}
\end{figure}

Figs.~\ref{fig:load} and \ref{fig:latency} carry the link model to the
transport level with 1500-B packets and stop-and-wait ARQ. Goodput
saturates at 93--96\% of the line rate (ACK overhead accounts for the
remainder), and 99th-percentile latency stays below 0.11~ms at 70\%
load at all demonstrated ranges---comfortably within the budgets of
interactive and industrial-control traffic, and the property that makes
the diffused-LD architecture interesting for low-latency optical
access. The ns-3 example programs reproduce these scenarios over the
full IPv4/UDP stack.

\section{What the Prototype Established, and What the Framework Adds}
\label{sec:loopholes}

\begin{table}[!t]
\caption{Prototype Gaps and Their Resolution in This Work}
\label{tab:loopholes}
\centering
\begin{tabular}{p{0.30\columnwidth}p{0.58\columnwidth}}
\toprule
Gap in the prototype study & Resolution\\
\midrule
127-Mb/s rate estimate from MATLAB timing & UART ceiling analysis
(Fig.~\ref{fig:uart}); rate claims re-derived from the validated link
budget (Fig.~\ref{fig:ratedist})\\
``No BER observed'' at 14~m & Noise-based BER-vs-distance curves
(Fig.~\ref{fig:berdist}); at UART rates the model predicts BER
$<10^{-12}$ at 14~m, consistent with observing zero errors\\
No noise/SNR model & Shot $+$ thermal budget
\eqref{eq:shot}--\eqref{eq:snr}, validated by Monte-Carlo and pinned by
unit tests\\
Single hardware operating point & Full design-space sweeps: distance,
power, beam width, modulation, coverage
(Figs.~\ref{fig:gain}--\ref{fig:coverage})\\
Eye safety asserted qualitatively & Diffused extended-source operation
discussed against IEC~60825-1~\cite{iec60825}; power--range trade
quantified (Fig.~\ref{fig:qfactor}b)\\
No network-level evidence & ns-3 full-stack simulations
(Figs.~\ref{fig:load}--\ref{fig:latency})\\
No reproducibility & Open ns-3 module $+$ Python engine, cross-validated
to 0.1\%\\
\bottomrule
\end{tabular}
\end{table}

Table~\ref{tab:loopholes} summarizes the forensic accounting. We
emphasize scope honestly: the hardware prototype \emph{established}
end-to-end feasibility of the diffused-LD architecture at 14~m,
including voice and image payloads; it \emph{could not establish}
multi-Mb/s throughput, error floors, or coverage. All higher-rate
figures in this paper are model-based predictions from a link budget
whose every component is standard
\cite{kahn1997wireless,komine2004fundamental,ghassemlooy2019optical},
whose parameters are prototype-derived, and whose implementations are
open and mutually validating. Experimental confirmation at full
bandwidth---a wideband driver and receiver replacing the UART
path---is the natural next step and is enabled, not replaced, by this
framework.

Two modeling limitations bound the claims. First, the channel is LOS
only; diffuse multipath adds a floor of delayed power that matters for
wide-FoV receivers near walls~\cite{kahn1997wireless,hussein2015mobile}
and will be added as an impulse-response extension of
\texttt{LifiOpticalChannelModel}. Second, the M-QAM subcarrier chain
assumes ideal DC bias and linear LD response; nonlinearity and clipping
studies (as in DCO-/ACO-OFDM analyses~\cite{islim2016modulation,
niaz2017total}) are orthogonal and composable with the framework.

\section{Conclusion}

We transformed a working but interface-limited LD-LiFi prototype into a
complete, reproducible system study. A diffuser-parameterized Lambertian
link model with a full noise budget, validated by Monte-Carlo simulation
and pinned across two independent implementations, shows that the
prototype's optical front end supports 930~Mb/s net at its demonstrated
14-m range and 1.86~Gb/s at short range under a standard HD-FEC
threshold, while OOK persists to 23~m. Coverage maps quantify the
diffuser's rate-versus-coverage dial, and full-stack ns-3 simulations
demonstrate near-line-rate goodput with sub-0.11-ms tail latency. The
open \texttt{lifi} ns-3 module makes laser-based LiFi accessible to
network-layer research---multi-cell interference, handover, hybrid
RF/optical scheduling---on a physically validated foundation. Future
work adds measured wideband hardware, diffuse multipath, LD
nonlinearity, and OFDM bit-loading.

\bibliographystyle{IEEEtran}
\bibliography{refs}

@article{chowdhury2018comparative,
  author  = {Chowdhury, M. Z. and Hossan, M. T. and Islam, A. and Jang, Y. M.},
  title   = {A Comparative Survey of Optical Wireless Technologies: Architectures and Applications},
  journal = {IEEE Access},
  volume  = {6},
  pages   = {9819--9840},
  year    = {2018}
}

@article{zafar2017laser,
  author  = {Zafar, F. and Bakaul, M. and Parthiban, R.},
  title   = {Laser-Diode-Based Visible Light Communication: Toward Gigabit Class Communication},
  journal = {IEEE Communications Magazine},
  volume  = {55},
  number  = {2},
  pages   = {144--151},
  year    = {2017}
}

@article{gao2018real,
  author  = {Gao, Y.-L. and Wu, Z.-Y. and Wang, Z.-K. and Wang, J.},
  title   = {A 1.34-{Gb/s} Real-Time {Li-Fi} Transceiver With {DFT}-Spread-Based {PAPR} Mitigation},
  journal = {IEEE Photonics Technology Letters},
  volume  = {30},
  number  = {16},
  pages   = {1447--1450},
  year    = {2018}
}

@article{islim2017towards,
  author  = {Islim, M. S. and Ferreira, R. X. and He, X. and Xie, E. and Videv, S. and Viola, S. and Watson, S. and Bamiedakis, N. and Penty, R. V. and White, I. H. and Kelly, A. E. and Gu, E. and Haas, H. and Dawson, M. D.},
  title   = {Towards 10 {Gb/s} Orthogonal Frequency Division Multiplexing-Based Visible Light Communication Using a {GaN} Violet Micro-{LED}},
  journal = {Photonics Research},
  volume  = {5},
  number  = {2},
  pages   = {A35--A43},
  year    = {2017}
}

@article{islim2016modulation,
  author  = {Islim, M. S. and Haas, H.},
  title   = {Modulation Techniques for {Li-Fi}},
  journal = {ZTE Communications},
  volume  = {14},
  number  = {2},
  pages   = {29--40},
  year    = {2016}
}

@article{tsonev2014gbs,
  author  = {Tsonev, D. and Chun, H. and Rajbhandari, S. and McKendry, J. J. D. and Videv, S. and Gu, E. and Haji, M. and Watson, S. and Kelly, A. E. and Faulkner, G. and Dawson, M. D. and Haas, H. and O'Brien, D.},
  title   = {A 3-{Gb/s} Single-{LED} {OFDM}-Based Wireless {VLC} Link Using a Gallium Nitride $\mu${LED}},
  journal = {IEEE Photonics Technology Letters},
  volume  = {26},
  number  = {7},
  pages   = {637--640},
  year    = {2014}
}

@article{shen2016underwater,
  author  = {Shen, C. and Guo, Y. and Oubei, H. M. and Ng, T. K. and Liu, G. and Park, K.-H. and Ho, K.-T. and Alouini, M.-S. and Ooi, B. S.},
  title   = {20-Meter Underwater Wireless Optical Communication Link with 1.5 {Gbps} Data Rate},
  journal = {Optics Express},
  volume  = {24},
  number  = {22},
  pages   = {25502--25509},
  year    = {2016}
}

@article{ghassemlooy2013visible,
  author  = {Haigh, P. A. and Ghassemlooy, Z. and Rajbhandari, S. and Papakonstantinou, I. and Popoola, W.},
  title   = {Visible Light Communications: 3.75 {Mbits/s} Data Rate with a 160 {kHz} Bandwidth Organic Photodetector and Artificial Neural Network Equalization},
  journal = {Photonics Research},
  volume  = {1},
  number  = {2},
  pages   = {65--68},
  year    = {2013}
}

@article{ho2018gigabit,
  author  = {Ho, K.-T. and Chen, R. and Liu, G. and Shen, C. and Holguin-Lerma, J. and Al-Saggaf, A. A. and Ng, T. K. and Alouini, M.-S. and He, J.-H. and Ooi, B. S.},
  title   = {3.2 Gigabit-per-Second Visible Light Communication Link with {InGaN/GaN} {MQW} Micro-Photodetector},
  journal = {Optics Express},
  volume  = {26},
  number  = {3},
  pages   = {3037--3045},
  year    = {2018}
}

@article{chi2015gan,
  author  = {Chi, Y.-C. and Hsieh, D.-H. and Tsai, C.-T. and Chen, H.-Y. and Kuo, H.-C. and Lin, G.-R.},
  title   = {450-nm {GaN} Laser Diode Enables High-Speed Visible Light Communication with 9-{Gbps} {QAM-OFDM}},
  journal = {Optics Express},
  volume  = {23},
  number  = {10},
  pages   = {13051--13059},
  year    = {2015}
}

@inproceedings{shen2017beyond,
  author    = {Shen, C. and Guo, Y. and Sun, X. and Liu, G. and Ho, K.-T. and Ng, T. K. and Alouini, M.-S. and Ooi, B. S.},
  title     = {Going Beyond 10-Meter, {Gbit/s} Underwater Optical Wireless Communication Links Based on Visible Lasers},
  booktitle = {Proc. Opto-Electronics and Communications Conference (OECC) and Photonics Global Conference (PGC)},
  year      = {2017}
}

@inproceedings{cogalan2015power,
  author    = {Cogalan, T. and Haas, H. and Panayirci, E.},
  title     = {Power Control-Based Multi-User {Li-Fi} Using a Compound Eye Transmitter},
  booktitle = {Proc. IEEE Global Communications Conference (GLOBECOM)},
  year      = {2015}
}

@article{niaz2017total,
  author  = {Niaz, M. T. and Imdad, F. and Kim, S. and Kim, H. S.},
  title   = {Total Least-Square-Based Receiver for Asymmetrically Clipped Optical-{OFDM} Visible Light Communication System},
  journal = {IET Optoelectronics},
  volume  = {11},
  number  = {4},
  pages   = {129--133},
  year    = {2017}
}

@article{wu2018white,
  author  = {Wu, T.-C. and Chi, Y.-C. and Wang, H.-Y. and Tsai, C.-T. and Huang, Y.-F. and Lin, G.-R.},
  title   = {White-Lighting Communication With a {Lu$_3$Al$_5$O$_{12}$:Ce$^{3+}$/CaAlSiN$_3$:Eu$^{2+}$} Glass Covered 450-nm {InGaN} Laser Diode},
  journal = {Journal of Lightwave Technology},
  volume  = {36},
  number  = {9},
  pages   = {1634--1643},
  year    = {2018}
}

@article{chi2017violet,
  author  = {Chi, Y.-C. and Huang, Y.-F. and Wu, T.-C. and Tsai, C.-T. and Chen, L.-Y. and Kuo, H.-C. and Lin, G.-R.},
  title   = {Violet Laser Diode Enables Lighting Communication},
  journal = {Scientific Reports},
  volume  = {7},
  number  = {1},
  pages   = {10469},
  year    = {2017}
}

@article{hussein2015mobile,
  author  = {Hussein, A. T. and Elmirghani, J. M. H.},
  title   = {Mobile Multi-Gigabit Visible Light Communication System in Realistic Indoor Environment},
  journal = {Journal of Lightwave Technology},
  volume  = {33},
  number  = {15},
  pages   = {3293--3307},
  year    = {2015}
}

@article{kahn1997wireless,
  author  = {Kahn, J. M. and Barry, J. R.},
  title   = {Wireless Infrared Communications},
  journal = {Proceedings of the IEEE},
  volume  = {85},
  number  = {2},
  pages   = {265--298},
  year    = {1997}
}

@article{komine2004fundamental,
  author  = {Komine, T. and Nakagawa, M.},
  title   = {Fundamental Analysis for Visible-Light Communication System Using {LED} Lights},
  journal = {IEEE Transactions on Consumer Electronics},
  volume  = {50},
  number  = {1},
  pages   = {100--107},
  year    = {2004}
}

@article{haas2016what,
  author  = {Haas, H. and Yin, L. and Wang, Y. and Chen, C.},
  title   = {What is {LiFi}?},
  journal = {Journal of Lightwave Technology},
  volume  = {34},
  number  = {6},
  pages   = {1533--1544},
  year    = {2016}
}

@book{ghassemlooy2019optical,
  author    = {Ghassemlooy, Z. and Popoola, W. and Rajbhandari, S.},
  title     = {Optical Wireless Communications: System and Channel Modelling with {MATLAB}},
  edition   = {2nd},
  publisher = {CRC Press},
  year      = {2019}
}

@misc{ieee802157,
  author       = {{IEEE}},
  title        = {{IEEE} Standard for Local and Metropolitan Area Networks---Part 15.7: Short-Range Optical Wireless Communications},
  howpublished = {IEEE Std 802.15.7-2018},
  year         = {2018}
}

@misc{iec60825,
  author       = {{IEC}},
  title        = {Safety of Laser Products---Part 1: Equipment Classification and Requirements},
  howpublished = {IEC 60825-1:2014},
  year         = {2014}
}

@incollection{riley2010ns3,
  author    = {Riley, G. F. and Henderson, T. R.},
  title     = {The ns-3 Network Simulator},
  booktitle = {Modeling and Tools for Network Simulation},
  publisher = {Springer},
  pages     = {15--34},
  year      = {2010}
}

@inproceedings{aldalbahi2016ns3,
  author    = {Aldalbahi, A. and Rahaim, M. and Khreishah, A. and Ayyash, M. and Ackerman, R. and Basuino, J. and Berreta, W. and Little, T. D. C.},
  title     = {Extending ns3 to Simulate Visible Light Communication at Network-Level},
  booktitle = {Proc. 23rd International Conference on Telecommunications (ICT)},
  year      = {2016}
}

@article{pathak2015visible,
  author  = {Pathak, P. H. and Feng, X. and Hu, P. and Mohapatra, P.},
  title   = {Visible Light Communication, Networking, and Sensing: A Survey, Potential and Challenges},
  journal = {IEEE Communications Surveys \& Tutorials},
  volume  = {17},
  number  = {4},
  pages   = {2047--2077},
  year    = {2015}
}

@article{karunatilaka2015led,
  author  = {Karunatilaka, D. and Zafar, F. and Kalavally, V. and Parthiban, R.},
  title   = {{LED} Based Indoor Visible Light Communications: State of the Art},
  journal = {IEEE Communications Surveys \& Tutorials},
  volume  = {17},
  number  = {3},
  pages   = {1649--1678},
  year    = {2015}
}

\end{document}